# Fast computation of the statistical significance test for spatio-temporal receptive field estimates obtained using spike-triggered averaging of binary pseudo-random sequences

Assoc. Prof. Dr. Murat Okatan[a,b]

[a] Istanbul Technical University, Informatics Institute, Istanbul, Türkiye,

[b] Istanbul Technical University, Artificial Intelligence and Data Engineering Department, Istanbul, Türkiye

okatan@itu.edu.tr, ORCID ID: 0000-0002-0064-6747

Article Type: Research Paper

**Abstract**

**Background:** Spatio-temporal receptive fields (STRF) of visual neurons are often estimated using spike-triggered averaging of binary pseudo-random stimulus sequences. The stimuli are visual displays that contain black and white pixels that flicker randomly at a fixed frame rate without any spatial or temporal correlation. The spike train of a visual neuron, such as a retinal ganglion cell, is recorded simultaneously with the stimulus presentation. The neuron's STRF is estimated by averaging the stimulus frames that coincide with spikes at fixed latencies. Recently, an exact analytical method for determining the statistical significance of the estimated value of the STRF pixels has been developed. Application of the method on spike trains collected from individual mouse retinal ganglion cells revealed that the time required to compute the test ranged from a couple of minutes to half a day for different neurons. **New method:** Here, this method is accelerated by using the Normal approximation to the null distribution of STRF pixel estimates. **Results:** The significance threshold and computation time obtained under the approximate distribution are examined systematically as a function of various input spike trains collected from individual mouse retinal ganglion cells. **Comparison with existing methods:** The accuracy and the time saved by the use of the approximate distribution are examined in comparison with the exact distribution. **Conclusions**: For the real data analyzed here, the approximate distribution yields the same significance thresholds as the exact distribution within a much shorter computation time.

## 1. Introduction

The study of neural receptive fields has been central to understanding how the nervous system functions (Hubel & Wiesel, 1968; Kuffler, 1953; Sherrington, 1906). In early studies, the term "receptive field" was used to refer to that patch of the skin surface, which, when touched, elicited a reflex action, such as the scratch-reflex (Sherrington, 1906). Reflexes were an indirect way of observing the firing of neural action potentials, or spikes. With the progress in neural recording methods, it has been possible to monitor the neural spike trains of individual neurons during light anesthesia and awake behavior (Hubel, 1957, 1958, 1959), which allowed mapping the stimulus subspace that modulates neural activity (Hubel & Wiesel, 1968, 1959; Marg et al., 1968). As a result, the term receptive field now refers not only to a patch of sensory surface, but to a mathematical construct that is defined in many-dimensional spaces, where the dimensions may include variables such as time, space, and spatial or temporal frequency (Bigelow & Malone, 2017; Chang et al., 2010; Lindeberg, 2016; Shih et al., 2020; Waleszczyk et al., 2007; Wallace et al., 1992; Wallace & Stein, 1996). For visual neurons, this construct usually characterizes a spatio-temporal receptive field (STRF) (Cai et al., 1997; Chichilnisky, 2001; Eckhorn et al., 1993; Jones & Palmer, 1987; Kim et al., 2008; Lefebvre et al., 2008; Li et al., 2018; Lindeberg, 2016; Malone et al., 2007; McLean & Palmer, 1989; Pamplona et al., 2022; Ringach et al., 1997a; Schwartz et al., 2006; Seilheimer et al., 2020; Sharpee, 2013; Theunissen et al., 2001; Yasui et al., 1979).

Various types of stimuli are used in estimating the STRF of visual cells, including finite sets of orthonormal stimuli (Malone et al., 2007; Ringach et al., 1997a), grating-like stimuli (Emerson et al., 1987; Waleszczyk et al., 2007), randomly moving single or collections of point sources (Eckhorn et al., 1993; McLean & Palmer, 1989), flashing a small bar of light at various locations (Stevens & Gerstein, 1976), and pseudo-random binary sequences, also called m-sequences, white noise or non-Gaussian white noise (Anzai et al., 1999b, 1999a; Cai et al., 1997; Citron et al., 1981; DeAngelis et al., 1993; Jacobson et al., 1993; Jones & Palmer, 1987; Li et al., 2018; Menz & Freeman, 2004; Reid & Shapley, 1992). Among these stimuli, pseudo-random binary sequences allow for the estimation of the STRF in a relatively much shorter amount of recording time, since stimulus presentation and response recording are continuous and not interrupted by inter-stimulus-intervals (DeAngelis et al., 1995).

The STRF is usually estimated through spike-triggered averaging (STA) of the binary pseudo-random stimulus sequences. STA consists of taking a snapshot of the stimulus at physiologically-relevant latencies, relative to spiking, every time the neuron generates a spike, and computing the average of those snapshots. STA, which is also referred to as "reverse correlation", reveals the average spatio-temporal stimulus history that precedes spiking, and is widely used in estimating the STRF (Bigelow & Malone, 2017; Cai et al., 1997; Chang et al., 2010; Chichilnisky, 2001; Eckhorn et al., 1993; Jones & Palmer, 1987; Kim et al., 2008; Lefebvre et al., 2008; Li et al., 2018; Malone et al., 2007; McLean et al., 1994; Pamplona et al., 2022; Ringach et al., 1997a; Schwartz et al., 2006; Seilheimer et al., 2020; Sharpee, 2013; Shih et al., 2020; Theunissen et al., 2001; Yasui et al., 1979).

In binary pseudo-random visual stimulus sequences, the stimulus consists of black and white pixels that flicker randomly at a fixed frame rate (Anzai et al., 1999b, 1999a; Brown et al., 2000; Cai et al., 1997; Citron et al., 1981; DeAngelis et al., 1993; Jacobson et al., 1993; Jones & Palmer, 1987; Kim et al., 2008; Lefebvre et al., 2008; Li et al., 2018; Meister et al., 1994; Menz & Freeman, 2004; Reid et al., 1997; Reid & Alonso, 1995; Reid & Shapley, 1992; Shah et al., 2020; Soo et al., 2011). In computing the STA, the stimulus value of each white pixel is taken as 1, while that of each black pixel is taken as -1. Thus, the estimated value of STRF pixels ranges from -1 to 1 at all pixels considered. Depending on whether the visual stimuli consist of lines (1-D) or checkerboard patterns (2-D), STRF estimates become 2-D or 3-D pixel maps, respectively, with time-before-spike as the additional dimension.

Although the structural and functional properties of STRF estimates are examined by extracting certain features, such as the spatial extent (functional cell size) (Cai et al., 1997; Lefebvre et al., 2008; Li et al., 2018; Malone et al., 2007; Moore et al., 2011; Ringach et al., 1997b; Seilheimer et al., 2020; Yasui et al., 1979), spatial symmetry, biphasic time course and orientation (Kim et al., 2008; Li et al., 2018), time to peak (Cai et al., 1997; Lefebvre et al., 2008; Li et al., 2018; Moore et al., 2011) and ratio of the integrated or peak amplitudes in the dark and light areas (Cai et al., 1997; Seilheimer et al., 2020), an exact analytical test for determining the statistical significance of the estimated STRF pixel values was lacking until recently. In a recent study, the exact distribution of the estimated STRF pixel values has been derived for binary pseudo-random sequences and given spike trains, under the null hypothesis that spike occurrences and stimulus values are statistically independent (Okatan, 2023a, 2023b). Here, this test is referred to as the "STA-BPRS test" for short, where the initialism "STA-BPRS" stands for "Spike-Triggered Average Binary Pseudo-Random Stimulus". It was shown that the computation of the STA-BPRS test took a few minutes to half a day, depending on the number of stimulus frames for which more than one spike was observed per frame (Okatan, 2023a). Here, further evidence is provided in new cells, indicating that this duration reaches even much longer periods of time for some cells. It is therefore of interest to accelerate the computation of the STA-BPRS test. Here, this acceleration is achieved by using a Normal approximation to the null distribution of STRF pixel estimates. The accuracy and the time saved by the use of the approximate distribution are examined systematically as a function of various input spike trains collected from individual mouse retinal ganglion cells (Zhang et al., 2014).

## 2. Materials and methods

### 2.1 Data and Computing

The data used here were downloaded from CRCNS.ORG (Zhang et al., 2014) and were originally collected from the retinas of mutant mice in experiments that investigated the role of gamma-protocadherins in the formation of functional retinal neural circuits (Lefebvre et al., 2008). During the recordings, a binary pseudo-random stimulus sequence was presented for approximately 25 min with a frame rate of 60 Hz. Data from two animals have been used here. Spike trains of 7 (34) retinal ganglion cells were available from animal 20080516_R1 (20080516_R2). The data of 20080516_R1 were previously analyzed in the study where the STA-BPRS test was initially developed (Okatan, 2023a); further details about the data are available therein.

All computations were performed under MATLAB (R2020b) on a Mobile Workstation with 16 GB ECC RAM and Intel® Xeon® W-10855M CPU @ 2.80 GHz, 6 Cores and 12 Logical Processors.

## 3. Theory and calculation

This section starts by presenting the formula for the estimated value of STRF pixels. Next, the formula for the probability mass function of this estimated value is provided for given spike trains, under the null hypothesis, $H_0$, that spiking is statistically independent from the stimulus presentation. After pointing out the reason why the computation of the null probability mass function (pmf) takes a very long time for some cells, the Normal approximation to the null pmf is explained. The section is concluded by explaining how the exact and the approximate distributions are compared against each other.

### 3.1 Estimated Value of the STRF Pixels

Let $s(x, u)$ denote the stimulus intensity at location $x$ and time $u$ in a binary pseudo-random sequence, where the probability of each binary value is 0.5. Here, $x$ is assumed to represent a one-dimensional discrete spatial coordinate, or pixel address, but it can represent a pixel address in a high dimensional stimulus space without any loss of generality. The value of $s(x, u)$ is taken to be either 1 or -1 (Kim et al., 2008; Lefebvre et al., 2008; Meister et al., 1994; Reid & Alonso, 1995; Reid & Shapley, 1992; Shah et al., 2020; Soo et al., 2011). It follows that $2^{-1}(1 + s(x, u)) \sim B(1, 0.5)$, where $B(1, 0.5)$ denotes the Binomial distribution with a success probability of 0.5 and number of trials of 1.

An illustration of $s(x, u)$ and coincident neural spikes is shown in Fig. 1 for 10 stimulus frames with a frame rate of $\tau^{-1} = 60$ Hz at some fixed spatial coordinate $x$, where $\tau$ is the frame duration. Frames 2, 5 and 8 have no spikes, frames 1, 7 and 10 have 1 spike, frames 3, 4 and 6 have 2, 3 and 4 spikes, respectively.

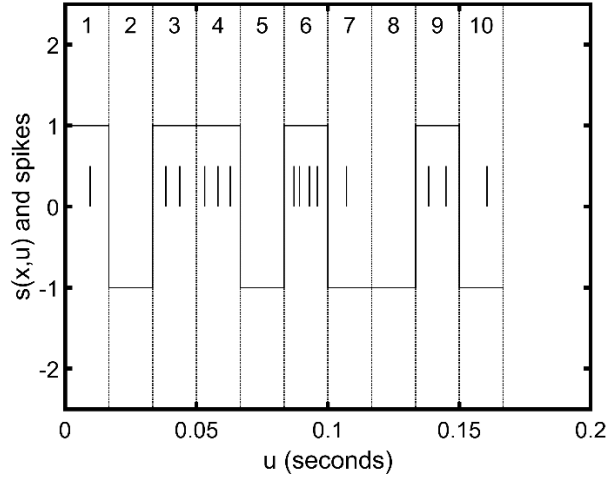

Figure 1. A hypothetical illustration of $s(x, u)$ and coincident neural spikes from a neuron. Frame numbers are shown at the top, from 1 to 10. Vertical line segments represent spikes.

The estimated value of the STRF pixel at location $x$ and time $t$ is given by Eq. 1 (Kim et al., 2008; Theunissen et al., 2001; Yasui et al., 1979):

$$h(x, t) = \frac{1}{n} \sum_{i=1}^{n} s(x, u_i + t), \qquad (1)$$

where, $n$ is the number of spikes used in constructing the STRF, $u_i$ is the time of the $i^{th}$ spike, and $t \leq 0$ is a latency between the stimulus values and the spike times. Negative $t$ values are used when the presentation of the stimulus precedes the observed spikes. In this way the STRF represents the average stimulus intensity that was observed at time $t$ relative to spike times, at location $x$.

If an STRF duration of $q$ frames is desired, then $-q\tau < t \leq 0$, and the spikes that occur during the first $q$ frames are ignored in Eq. 1. For instance, if $q = 2$, then, in Fig. 1, the first spike would not be considered in the computation of Eq. 1.

### 3.2 The PMF of $h(x, t)$ under $H_0$

The derivation of the null pmf of $h(x,t)$, denoted by $f_h^J(h(x,t)|H_0, n_{1:J})$, is explained in detail in (Okatan, 2023a). Here, the formula of this pmf is presented before its approximate form is introduced.

In a binary pseudo-random stimulus sequence, each stimulus frame is presented for a fixed duration, such as $\tau =16.67$ ms (60 Hz) (Kim et al., 2008; Lefebvre et al., 2008). A neuron may fire more than one spike during that period, as illustrated in Fig. 1. Let $J$ denote the maximum number of spikes fired by a neuron during any stimulus frame. Then, the total number of spikes fired by the neuron during the recording can be expressed as $n = \sum_{j=1}^{J} j n_j$, where $n_j$ denotes the number of stimulus frames during which the neuron fires exactly $j$ spikes. For the brief recording illustrated in Fig. 1, if $q = 2$, then $J = 4$, $n_1 = 2$, $n_2 = 2$, $n_3 = 1$, $n_4 = 1$, and $n = 13$. Representing the set of numbers $n_1, n_2, …, n_J$ by $n_{1:J}$, $f_h^J(h(x,t)|H_0, n_{1:J})$ is expressed as:

$$f_h^J(h(x,t)|H_0, n_{1:J}) = \\ \sum_{m_1=-n_1:2:n_1} \sum_{m_2=-n_2:2:n_2} \cdots \sum_{m_J=-n_J:2:n_J} \mathbb{1}\left(nh(x,t) == \sum_{j=1}^{J} j m_j\right) \prod_{j=1}^{J} f_S(m_j|H_0, n_j), \quad (2)$$

where

$$f_S(m_j|H_0, n_j) = (1 + (-1)^{n_j+m_j}) \binom{n_j}{\frac{n_j+m_j}{2}} 0.5^{n_j+1}, \quad (3)$$

with $-n_j \leq m_j \leq n_j$ (Okatan, 2023a).

In Eq. 2, the indicator function $\mathbb{1}(\cdot)$ has a value of 1 if its argument is true, and 0 otherwise. Note that each $m_j$ increments by 2 because $f_S(m_j|H_0, n_j)$ is zero at skipped values per Eq. 3. Because $f_S(-m_j|H_0, n_j) = f_S(m_j|H_0, n_j)$ for $j = 1:J$, both $f_S(m_j|H_0, n_j)$ and $f_h^J(h(x,t)|H_0, n_{1:J})$ are symmetrical about 0, as expected.

### 3.3 The Computation of the Null PMF Takes a Very Long Time for Some Cells

The computation of Eq. 2 may take a prohibitively long time if $n_j$ is on the order of $10^3$ for several $j$, which may occasionally happen for some neurons. That's because Eq. 2 involves the summation of $v_{1:J} = \prod_{j=1}^{J}(n_j + 1)$ terms. This product ranges from 5 $10^3$ to 5 $10^{15}$ in different cells (Table A.1). The next section explains that the null pmf can be computed as a combination of the exact and the approximate distributions by approximating the product of $f_S(m_j|H_0, n_j)$ in Eq. 2 by a Normal distribution for large $n_j$.

### 3.4 Normal Approximation to the Null PMF

Let $S_n$ denote the quantity $nh(x,t)$ that appears in Eq. 2. According to Eq. 1, $S_n = \sum_{i=1}^{n} s(x, u_i + t)$, which is the spike-triggered sum of the stimulus values. The null pmf of $S_n$ is (Okatan, 2023a)

$$f_S^J(S_n|H_0, n_{1:J}) = \sum_{m_1=-n_1:2:n_1} \sum_{m_2=-n_2:2:n_2} \cdots \sum_{m_J=-n_J:2:n_J} \mathbb{1}(S_n == \sum_{j=1}^{J} jm_j) \prod_{j=1}^{J} f_S(m_j|H_0, n_j), \quad (4)$$

which is related to Eq. 2 through the relation $f_S^J(S_n|H_0, n_{1:J}) = f_h^J(n^{-1}S_n|H_0, n_{1:J})$.

$S_n$ can be partitioned into two sums that are contributed by frames where $n_j > T$ and $n_j \leq T$, for a relatively large number $T$ to be determined below. This partitioning may be useful because the approximation works better when $n_j$ is large, as shown below, and thus only $f_S(m_j|H_0, n_j)$ with large $n_j$ may be targeted for the approximation. Therefore, the goal of the following analysis is to use an approximate distribution for the component of $S_n$ that is due to frames with large $n_j$, while preserving the exact distributions for the remaining components of $S_n$.

To determine $T$, let $G$ and $\tilde{G}$ denote the sets of $j$ for which $n_j \leq T$ and $n_j > T$, respectively. Then, $S_n = S_G + S_{\tilde{G}}$, where $S_G = \sum_{j \in G} S_{n_j}$ and $S_{\tilde{G}} = S_n - S_G = \sum_{j \in \tilde{G}} S_{n_j}$. In these sums, $S_{n_j}$ represents the spike-triggered stimulus sum contributed by the $n_j$ frames that contain $j$ spikes each.

Per the Central Limit Theorem (Liapounov) for the sum of independent random variables (DeGroot & Schervish, 2011), the distribution of $S_{\tilde{G}}$ is approximately Normal with mean $\mu_{\tilde{G}} = \sum_{j \in \tilde{G}} E(S_{n_j})$ and variance $\sigma_{\tilde{G}}^2 = \sum_{j \in \tilde{G}} Var(S_{n_j})$. The Normal approximation allows for computing an approximate value for the probability of $S_{\tilde{G}}$ in one step, instead of adding on the order of $\prod_{j \in \tilde{G}}(n_j + 1)$ terms within Eq. 2 to compute it exactly.

The computation of the probability of $S_G$ requires adding on the order of $v_G = \prod_{j \in G}(n_j + 1)$ terms. Since the purpose of using the approximation is to reduce the computation time to an acceptable level, it would make sense to keep the cardinality of $G$ low, which could be achieved by choosing $T$ as small as possible. On the other hand, keeping $T$ high is desirable for minimizing the adverse effects of using approximate distributions. Since $n_j$ tends to decrease quickly with increasing $j$ in real spike trains (Table A.1), the cardinality of $G$ may remain sufficiently low even when $T$ is chosen relatively large. For instance, if the time taken to add $10^6$ numbers takes a tolerable amount of time on a computer, and if $v_G$ remains below $10^6$ when $T \leq 1000$ for a neuron, then $T = 1000$ could be used for that neuron.

The foregoing implies that a maximum allowable number of terms to be added may be preset for the computing system at hand — let $\Omega$ denote that number — and the maximum value of $T$ can be determined for each neuron, such that $v_G < \Omega$. In other words, denoting the $k^{th}$ order statistic of $n_1, n_2, \ldots, n_J$ by $n_{(k)}$,

$$T = \max_{1 \leq k \leq J}\{n_{(k)} | \prod_{j=1}^{k}(n_{(j)} + 1) < \Omega\}, \quad (5)$$

may be used. If the inequality in Eq. 5 cannot be satisfied even at $k = 1$, then $T = 0$ may be set to select all $n_j$ for approximation.

Once $\tilde{G}$ is determined, since $E\left(S_{n_j}\right) = 0$ and $Var\left(S_{n_j}\right) = j^2 n_j$, we have $S_{\tilde{G}}$ distributed approximately $N(0, \sigma_{\tilde{G}}^2)$, where $\sigma_{\tilde{G}}^2 = \sum_{j \in \tilde{G}} j^2 n_j$. Thus, $\Phi(\cdot)$ denoting the standard Normal cumulative distribution function,

$$Pr(S_{\tilde{G}} = m | H_0, n_{j \in \tilde{G}}) \approx \mathbb{1}\left(|m| \leq \sum_{j \in \tilde{G}} jn_j\right)\left\{\Phi\left(\frac{m+1}{\sigma_{\tilde{G}}}\right) - \Phi\left(\frac{m-1}{\sigma_{\tilde{G}}}\right)\right\} = \mathfrak{f}_S^{\tilde{G}}(m | H_0, n_{j \in \tilde{G}}), \quad (6)$$

which is corrected for continuity (DeGroot & Schervish, 2011).

It follows that

$$f_S^J(S_n | H_0, n_{1:J}) \approx \mathfrak{f}_S^J(S_n | H_0, n_{1:J}) =$$
$$\sum_{m_{j_1} = -n_{j_1}:2:n_{j_1}} \sum_{m_{j_2} = -n_{j_2}:2:n_{j_2}} \cdots \sum_{m_{j_K} = -n_{j_K}:2:n_{j_K}} \sum_{m = -n_{\tilde{G}}:2:n_{\tilde{G}}}$$
$$\mathbb{1}\left(S_n == m + \sum_{k=1}^{K} j_k m_{j_k}\right) \mathfrak{f}_S^{\tilde{G}}(m | H_0, n_{j \in \tilde{G}}) \prod_{k=1}^{K} f_S(m_{j_k} | H_0, n_{j_k}), \quad (7)$$

where $j_k \in G$, $K$ is the cardinality of $G$, and $n_{\tilde{G}} = \sum_{j \in \tilde{G}} jn_j$. In the foregoing, the symbols $f$ and $\mathfrak{f}$ refer to exact and approximate distributions, respectively.

By defining

$$f_S^G(S_G | H_0, n_{j \in G}) =$$
$$\sum_{m_{j_1} = -n_{j_1}:2:n_{j_1}} \sum_{m_{j_2} = -n_{j_2}:2:n_{j_2}} \cdots \sum_{m_{j_K} = -n_{j_K}:2:n_{j_K}} \mathbb{1}\left(S_G == \sum_{k=1}^{K} j_k m_{j_k}\right) \prod_{k=1}^{K} f_S(m_{j_k} | H_0, n_{j_k}), \quad (8)$$

we obtain

$$\mathfrak{f}_S^J(S_n | H_0, n_{1:J}) = \sum_{m = -n_{\tilde{G}}:2:n_{\tilde{G}}} \mathfrak{f}_S^{\tilde{G}}(m | H_0, n_{j \in \tilde{G}}) f_S^G(S_n - m | H_0, n_{j \in G}). \quad (9)$$

This form of $\mathfrak{f}_S^J(S_n | H_0, n_{1:J})$ reveals its construction as the convolution of the approximate distribution of $S_{\tilde{G}}$ and the exact distribution of $S_G$. Since $S_n = S_G + S_{\tilde{G}}$, and since $S_G$ and $S_{\tilde{G}}$ are independent under $H_0$ given the spike train, we have

$$f_S^J(S_n|H_0, n_{1:J}) = \sum_{m=-n_{\tilde{G}}:2:n_{\tilde{G}}} f_S^{\tilde{G}}(m|H_0, n_{j\in\tilde{G}}) f_S^G(S_n - m|H_0, n_{j\in G}), \quad (10)$$

and Eq. 9 is only a re-statement of this formula with the approximate distributions $\mathfrak{f}_S^J$ and $\mathfrak{f}_S^{\tilde{G}}$ being used in place of the exact distributions $f_S^J$ and $f_S^{\tilde{G}}$, respectively.

Despite its utility in elucidating how $\mathfrak{f}_S^J$ is obtained from $\mathfrak{f}_S^{\tilde{G}}$ and $f_S^G$, Eq. 9 hides from view an important simplification that can be implemented in the computation of $\mathfrak{f}_S^J$. This simplification is visible in Eq. 7, where the summation indexed by $m$ does none other than sifting $m = S_n - \sum_{k=1}^K j_k m_{j_k}$ if $(S_n - \sum_{k=1}^K j_k m_{j_k}) + n_{\tilde{G}}$ is even. Thus, a more compact expression of Eq. 7 would be

$$\mathfrak{f}_S^J(S_n|H_0, n_{1:J}) = \sum_{m_{j_1}=-n_{j_1}:2:n_{j_1}} \sum_{m_{j_2}=-n_{j_2}:2:n_{j_2}} \cdots \sum_{m_{j_K}=-n_{j_K}:2:n_{j_K}}$$
$$\frac{1 + (-1)^{S_n + n_{\tilde{G}} - \sum_{k=1}^K j_k m_{j_k}}}{2} \mathfrak{f}_S^{\tilde{G}}(S_n - \sum_{k=1}^K j_k m_{j_k}|H_0, n_{j\in\tilde{G}}) \prod_{k=1}^K f_S(m_{j_k}|H_0, n_{j_k}). \quad (11)$$

This form of the equation clearly shows that there are at most $\nu_G = \prod_{j\in G}(n_j + 1)$ terms to be added in computing $\mathfrak{f}_S^J(S_n|H_0, n_{1:J})$ for a given $S_n$.

Finally, $\mathfrak{f}_h^J(h(x,t)|H_0, n_{1:J})$ is obtained as $\mathfrak{f}_S^J(nh(x,t)|H_0, n_{1:J})$:

$$\mathfrak{f}_h^J(h(x,t)|H_0, n_{1:J}) = \sum_{m_{j_1}=-n_{j_1}:2:n_{j_1}} \sum_{m_{j_2}=-n_{j_2}:2:n_{j_2}} \cdots \sum_{m_{j_K}=-n_{j_K}:2:n_{j_K}}$$
$$\frac{1 + (-1)^{nh(x,t) + n_{\tilde{G}} - \sum_{k=1}^K j_k m_{j_k}}}{2} \mathfrak{f}_S^{\tilde{G}}(nh(x,t) - \sum_{k=1}^K j_k m_{j_k}|H_0, n_{j\in\tilde{G}}) \prod_{k=1}^K f_S(m_{j_k}|H_0, n_{j_k}). \quad (12)$$

A particular case of this approximate distribution is obtained when the set $G$ is empty, which may be obtained when $\Omega$ is set very low. In that case we have $\tilde{G} = \{1, 2, \ldots, J\}$ and

$$\mathfrak{f}_h^J(h(x,t)|H_0, n_{1:J}) = \frac{1 + (-1)^{n(1+h(x,t))}}{2} \mathfrak{f}_S^{\tilde{G}}(nh(x,t)|H_0, n_{1:J}), \quad (13)$$

where $\mathfrak{f}_S^{\tilde{G}}(nh(x,t)|H_0, n_{1:J})$ is obtained from Eq. 6 using $\sigma_{\tilde{G}}^2 = \sum_{j=1}^J j^2 n_j$.

### 3.5 The Comparison of the Exact and the Approximate Distributions for $J = 1$

The computation of the null pmf of $h(x,t)$ does not take an impractically long time for any realistically large $n$ if $J = 1$ (Section 4.1). Thus, it is not necessary to use the approximation for $J = 1$. However, the approximate distribution is used here to illustrate how the difference between the exact and the approximate distributions changes with $n$ in this simple case. From Eq. 6 and Eq. 7

$$\mathfrak{f}_S^1(S_n|H_0, n_1) = \mathfrak{f}_S^1(S_n|H_0, n) = \mathfrak{f}_S^{\tilde{G}}(S_n|H_0, n) = \mathbb{1}(|S_n| \leq n)\left\{\Phi\left(\frac{S_n + 1}{\sqrt{n}}\right) - \Phi\left(\frac{S_n - 1}{\sqrt{n}}\right)\right\}, \quad (14)$$

The exact distribution of $S_n$ is (Okatan, 2023a)

$$f_S(S_n|H_0, n) = (1 + (-1)^{n+S_n})\binom{n}{\frac{n+S_n}{2}} 0.5^{n+1}. \quad (15)$$

which was used above in Eq. 3. Let $\mathfrak{f}_S(S_n|H_0, n)$ denote $\mathfrak{f}_S^1(S_n|H_0, n_1)$ for simplicity.

The maximum absolute difference $\Delta$ between $f_S(S_n|H_0, n)$ and $\mathfrak{f}_S(S_n|H_0, n)$ for $-n \leq S_n \leq n$ is used for assessing how well the approximation matches the exact distribution as a function of $n$. While $\Delta$ may be used as an overall measure of accuracy, it doesn't provide information about whether the two distributions select the same set of significant pixels when used in the statistical significance test. To address that question directly, let $h^{*-}(x,t)$ and $\tilde{h}^{*-}(x,t)$ denote the least significant negative pixel value that is below the inclusive 2.5th percentile of the exact and the approximate distributions, respectively. We have

$$h^{*-}(x,t) = \max_{-n \leq M \leq n}\{M | Mn^{-1} < \theta^-\}, \quad (16)$$
$$\tilde{h}^{*-}(x,t) = \max_{-n \leq M \leq n}\{M | Mn^{-1} < \tilde{\theta}^-\}, \quad (17)$$

where $\theta^-$ is the $\left(100\frac{\alpha}{2}\right)^{th}$ percentile (inclusive) of $f_h^J(h(x,t)|H_0, n_{1:J})$, and $\tilde{\theta}^-$ is its approximate counterpart:

$$\tilde{\theta}^- = n^{-1} \min_{-n \leq M \leq n}\left\{M \middle| \Phi\left(\frac{M+1}{\sqrt{n}}\right) \geq \frac{\alpha}{2}\right\}, \quad (18)$$

which is corrected for continuity (DeGroot & Schervish, 2011).

The time it takes to compute the test using the exact versus the approximate distributions is measured at $n = 10^5$ in 10 independent measurements and the results are tested for equal medians using the Wilcoxon signed rank test. The computation times are also measured for $n = 1,2,\ldots,10^5$ to examine their dependence on $n$ using simple linear regression.

### 3.6 The Comparison of the Exact and the Approximate Distributions for $J > 1$

To compare the exact and the approximate distributions for $J > 1$, the distributions and the amplitude thresholds they dictate were computed for the spike trains of real neurons for which the exact distribution could be computed within up to a few days. The maximum absolute difference ($\Delta$) between $f_S^J(S_n|H_0, n_{1:J})$ and $\tilde{f}_S^J(S_n|H_0, n_{1:J})$, along with $h^{*-}(x,t)$ and $\tilde{h}^{*-}(x,t)$, were computed and examined as a function of $n$. $\tilde{h}^{*-}(x,t)$ was computed using Eq. 17, where the $\left(100\frac{\alpha}{2}\right)^{th}$ percentile (inclusive) of $\tilde{f}_S^J(S_n|H_0, n_{1:J})$ was used as $\tilde{\theta}^-$.

The computation times of the test using the exact and the approximate distributions were measured for different values of $\Omega$. The non-parametric Kruskal-Wallis (KW) test is used for analyzing the time taken to compute the test for each animal and $\Omega$ value analyzed (Animals 20080516_R1/20080516_R2 × $\Omega$ = $\{1, 10^2, 10^4, 10^6, \text{Inf}\}$), supplemented by Tukey-Kramer multiple comparison test.

## 4. Results

This section first presents the results of comparing the exact and the approximate distributions for $J = 1$ in terms of the dependence of $\Delta$, $h^{*-}(x,t)$ and $\tilde{h}^{*-}(x,t)$ on $n$, and the time it takes to compute the tests. Next, the same comparisons are performed for $J > 1$.

### 4.1. The Comparison of the Exact and the Approximate Distributions for $J = 1$

For $J = 1$, both the exact and the approximate distributions are straightforward to compute, as explained in Section 3.5. Therefore, it is possible to compute them for very large numbers of spikes. Here, they are computed for $n = 1, 2, .., 10^5$. It is observed that the rate of change of $log_{10}(\Delta)$ with $log_{10}(n)$ is constant at large $n$, indicating that the approximation improves with increasing $n$, as expected (Fig. 2A). Simple linear regression analysis estimates that rate of change as -0.67 with a 95% confidence interval of -0.67 ± 1.33 10$^{-5}$.

To check whether the exact and the approximate distributions identify the same set of significant pixels, $h^{*-}(x,t)$ and $\tilde{h}^{*-}(x,t)$ are computed, which exist for $n > 5$. For $5 < n \leq 10^5$, the difference $h^{*-}(x,t) - \tilde{h}^{*-}(x,t)$ is identically equal to 0 for 99.757% of the $n$ values, and is identically equal to $2/n$ in the remaining 243 ('nonconforming') $n$ values (Fig. 2B). Note that $2/n$ is the smallest non-zero difference that can be observed between $h^{*-}(x,t)$ and $\tilde{h}^{*-}(x,t)$. The fact that the difference is never negative indicates that the approximate distribution is more conservative than the exact distribution at the nonconforming $n$ values considered here, and identifies the same set of pixels as the exact distribution for the remaining $n$ values. The difference of $2/n$ is observed even at large $n$, although its frequency of occurrence decreases with increasing $n$ (Fig. 2B). The set of nonconforming $n$ values shown in Fig. 2B were obtained for a two-tailed significance level of $\alpha = 0.05$; different sets of nonconforming $n$ values would be obtained for different significance levels used.

For $n = 10^5$, the computation of the exact threshold ($\theta^-$) took longer than the computation of the approximate threshold ($\tilde{\theta}^-$) in 10 independent measurements (83.7 ± 1.6 ms vs. 1.8 ± 0.2 ms, mean ± s.e.m.; p=1.95 10$^{-3}$ (two-tailed), Wilcoxon signed rank test). The computation times were shorter for smaller $n$. Simple linear regression estimated that the computation time of the exact threshold increases by 314.1 ns per spike, with a 95% confidence interval of [313.9 314.3] ns, whereas these figures are 9.57 ns and [9.56 9.58] ns per spike for the approximate threshold.

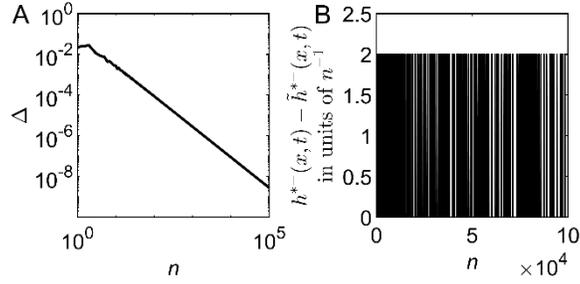

Figure 2. The effects of using $\tilde{f}_S(S_n|H_0, n)$ instead of $f_S(S_n|H_0, n)$. (A) The maximum absolute difference ($\Delta$) between the exact and the approximate distributions as a function of $n$. (B) The difference $h^{*-}(x,t) - \tilde{h}^{*-}(x,t)$ is either 0 or $2/n$ (see text for details).

## 4.2. The Comparison of the Exact and the Approximate Distributions for $J > 1$

For $J > 1$, the approximate distribution can be computed for different values of $\Omega$, as explained in Section 3.6. Here, results are shown for $\Omega = 1, 10^2, 10^4, 10^6$, and Inf (i.e. $\infty$), where $\Omega = $ Inf corresponds to the exact distribution, and $\Omega = 1$ results in the fully approximated distribution shown in Eq. 13.

The time taken to compute the test using the exact distribution increases with $\nu_{1:J}$ (Fig. 3). It took up to a few days to compute the tests for all but one of the 41 cells analyzed here; the computation of the exact test could not be completed for the remaining cell, which was cell 23 of animal 20080516_R2. The available measurements indicate that the logarithm of the computation time is a quadratic function of the logarithm of $\nu_{1:J}$. Based on the curve fit, the expected computation time for cell 23 is about 528 years (asterisk in Fig. 3). For that reason, $\Delta$ and $h^{*-}(x,t)$ were computed only for the remaining cells. For cell 23, the measured computation times for $\Omega = 1, 10^2, 10^4, 10^6$, are 242.8 ± 2.1 ms, 244.1 ± 2.2 ms, 406.4 ± 4.5 ms, and 113.5 ± 0.44 s, respectively (mean ± s.e.m for 100 independent measurements).

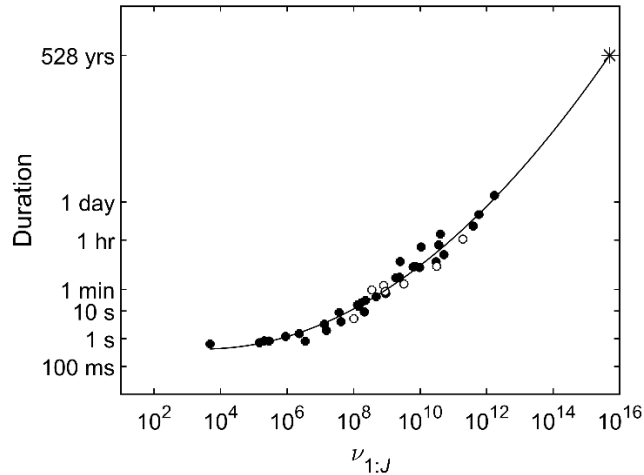

Figure 3. The time taken to compute the STA-BPRS test using the exact distribution (Eq. 2), which corresponds to $\Omega = $ Inf. The unfilled and filled circles correspond to the measurements obtained from the cells of animals 20080516_R1 and 20080516_R2, respectively. The expected computation time of cell 23 of animal 20080516_R2 is indicated by the asterisk.

The KW test reveals in which animal the duration changes significantly with $\Omega$. Cell 23 was excluded from this analysis due to the unavailability of the computation time measurement for $\Omega = \text{Inf}$. The KW test shows that the combination of animal and $\Omega$ value significantly affects the duration (H(9)=146.08, p=5.7e-27). Post-hoc Tukey-Kramer tests are used to compare all animal pairs of groups. No significant effects were found across animals (p>0.05). Therefore, the data from the two animals were pooled and the non-parametric KW test was used again for analyzing the time taken to compute the test for each $\Omega$ value analyzed, supplemented by Tukey-Kramer multiple comparison test. The second KW test shows that $\Omega$ value significantly affects the duration (H(4)=144.28, p=3.41e-30). Post-hoc Tukey-Kramer tests are used to compare durations in all $\Omega$ pairs. Significant effects were found in all $\Omega$ pairs but $\Omega = \{10^0, 10^2\}$, $\Omega = \{10^2, 10^4\}$ and $\Omega = \{10^6, \text{Inf}\}$ (Fig. 4).

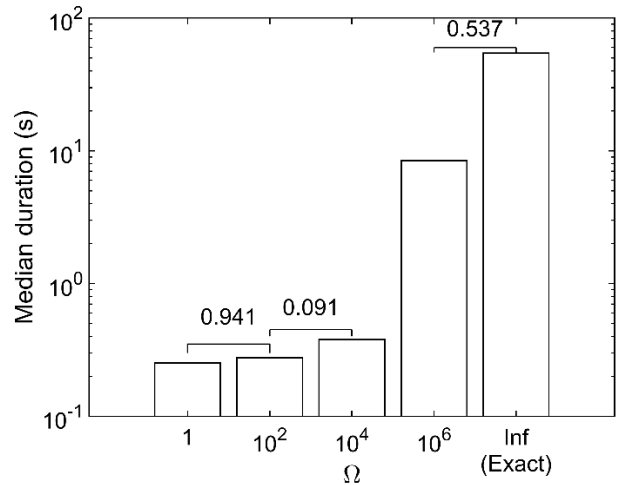

Figure 4. The time taken to compute the STA-BPRS test using various $\Omega$ values. Median duration was not different between the indicated pairs of bars. All the other bar pairs were significantly different (p<9.68 10$^{-3}$; post-hoc Tukey-Kramer test).

### 4.2.1 Dependence of $\Delta$ on $n$

As in the $J = 1$ case (Fig. 2A), the maximum absolute difference ($\Delta$) between the exact and the approximate distributions decreases with $n$ in the $J > 1$ case (Fig. 5). Also, $\Delta$ is observed to decrease with $\Omega$, as expected. Cell 23 was excluded from this analysis due to the unavailability of the exact distribution. Because $v_{1:J} < 10^4$ for cell 5 and $v_{1:J} < 10^6$ for cells 6, 15, 20 and 29 of 20080516_R2 (Table A.1), the exact distribution is obtained at $\Omega \geq 10^4$ or $\Omega = 10^6$ for those cells, which makes $\Delta = 0$ and $log_{10}(\Delta) = -\text{Inf}$. As a result, no marker could be plotted in Fig. 5 at those $\Omega$ values for those cells. For some other cells, markers were superimposed completely or to a large extent, hampering their visibility.

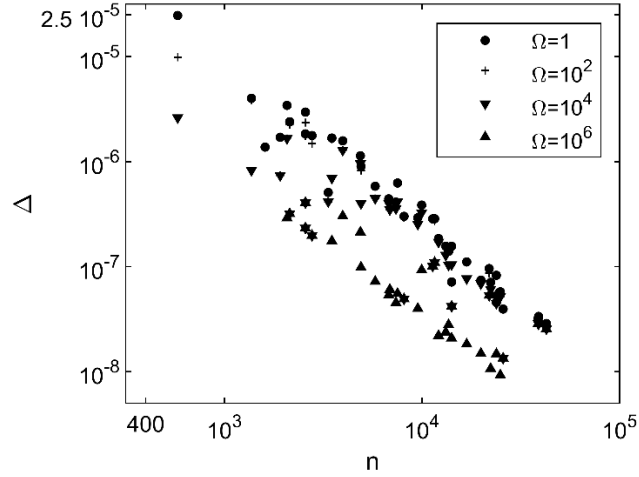

Figure 5. Dependence of $\Delta$ on $n$ and $\Omega$.

### 4.2.2 Exact and Approximate Distributions Identified the Same Set of Significant Pixels

Both the exact and the approximate distributions detected the same set of significant pixels. Namely, the difference $h^{*-}(x,t) - \tilde{h}^{*-}(x,t)$ was identically equal to 0 for all cells at all $\Omega$ values. By contrast, in the $J = 1$ case this difference was $2/n$ at some $n$ values, which were termed 'nonconforming'.

### 5. Discussion

Spike-triggered averaging of binary pseudo-random stimuli is a fast and relatively straightforward method for estimating the STRF of visual neurons (DeAngelis et al., 1995). Recently, a new statistical significance test has been proposed for identifying the significant pixels of such STRF estimates (Okatan, 2023a). Here, this test is referred to as the STA-BPRS test for short.

It was previously reported that the STA-BPRS test took up to 12 h to complete for cell 3 of animal 20080516_R1 (Okatan, 2023a). Here, after some code improvements, such as avoiding the use of global variables in recursive function calls, this duration was reduced down to about an hour for that cell (the rightmost open circle in Fig. 3). However, even with the new code, the test took about a day to complete for some cells of animal 20080516_R2, and could not be completed despite several days of execution for cell 23 of that animal. These long computation times prompted the development of the approximate null distributions presented here.

Because the null distribution is based on symmetrical Binomial distributions (Eq. 2 and Eq. 3), it can be well approximated by the Normal distribution with the same mean and variance (DeGroot & Schervish, 2011). The accuracy of the approximation may be kept high if only those components of the null distribution that are associated with large $n_j$ values are approximated, while the other components are kept exact. The selection of the pmf components on the basis of the magnitude of $n_j$ has been implemented here by the introduction of the $\Omega$ variable. $\Omega$ represents the maximum number of terms that the user is willing to add iteratively to compute the exact components of the pmf. Setting $\Omega$ too low (e.g. $\Omega = 1$) forces all components of the pmf to be approximated by the Normal distribution, whereas setting $\Omega$ too high (e.g. $\Omega = \text{Inf}$) avoids the approximation altogether and computes the exact pmf. In this way, it is possible to trade computation speed versus accuracy using a single variable.

The comparison of the exact and the approximate distributions has been performed first for $J = 1$ (Sections 3.5 and 4.1). This condition represents the situation where at most one spike is observed per stimulus frame throughout the experiment. It is found that, under this condition, the approximate distribution gets asymptotically closer to the exact distribution as the number of spikes, $n$, increases (Fig. 2A). Moreover, the significant pixels determined using the approximate distribution were identically equal to the significant pixels determined using the exact distribution at the large majority (99.757%) of $n$ values tested (Fig. 2B). In a small number of $n$ values — 243 to be exact — the least significant pixel value below the inclusive 2.5$^{th}$ percentile of the approximate distribution was smaller than the least significant pixel value below the inclusive 2.5$^{th}$ percentile of the exact distribution, by an amount that was equal to $2/n$, which is the smallest strictly positive difference that can exist between consecutive pixel values. The $n$ values where this difference was observed were referred to as 'nonconforming'.

It is highly unlikely that $J$ will be equal to 1 in a typical experiment, especially if firing rates are above a few Hz. The probability of $J = 1$ can be quantified in a hypothetical experiment, where the frame rate of the stimulus sequence is 60 Hz and the recorded spike train is a homogeneous Poisson process with a rate of $\phi$ Hz. In such a thought experiment, the number of spikes observed per frame would be distributed Poisson with mean $\phi/60$, placing the probability of at most 1 spike to be observed for a given frame to $p_{(0 \cup 1)} = p[0] + p[1] = e^{-\phi/60}(1 + \phi/60)$. An experimental record of duration $W$ seconds would contain $60W$ frames, and the probability of all of them containing at most 1 spike would be $\left(p_{(0 \cup 1)}\right)^{60W} = e^{-W\phi}(1 + \phi/60)^{60W}$. The recording duration for the data examined here was about 25 minutes, making $W \approx 1500\ s$. Thus, the highest firing rate that would yield $J = 1$ with probability more than 5% satisfies the inequality $-W\phi + 60W\ln(1 + \phi/60) > \ln(0.05)$, which yields $\phi \lesssim 0.5\ Hz$. This suggests that it is improbable to observe $J = 1$ for firing rates above 0.5 Hz during typical experiments. When $J = 1$ is observed, however, Fig. 2B suggests that it is better to use the exact distribution, rather than the approximate one, to avoid detecting false negative pixels in the rare instances when $n$ corresponds to a nonconforming value. Luckily, the exact distribution can be computed very quickly when $J = 1$.

The foregoing implies that $J$ will be larger than 1 in a typical experiment, and for some cells, as shown in the present data, the computation of the exact distribution will take impractically long times. In those cases, the approximate pmf developed here would have to be used to complete the test. In the present data the exact and the approximate tests detected the same set of significant pixels for 40 cells at all $\Omega$ values. It is harder to investigate the difference between the exact and the approximate distributions systematically using simulated data in the $J > 1$ case, compared to the $J = 1$ case, since several possible combinations of $n_j$ values would need to be considered for a given $n$. If such an exhaustive analysis were to be performed, perhaps 'nonconforming' $\tilde{G}$ sets could be identified. If the results obtained in the $J = 1$ case are any indication, the number of such nonconforming $\tilde{G}$ sets, if any, may constitute only a small fraction of the total number of possible $\tilde{G}$ sets. That may explain why the exact and the approximate distributions identified the same set of significant pixels in all cases where this comparison was possible (Section 4.2.2).

The results of the present study suggest the following course of action in analyzing large sets of cells. First, the quantity $v_{1:J}$ should be determined for all cells, as in Table A.1. Second, the cells should be sorted in the order of increasing $v_{1:J}$ and the STA-BPRS test should be performed in that order using $\Omega = \text{Inf}$. If the test cannot be completed for some cells after a time period that is deemed too long, then $\Omega$ should be reduced to the level of the largest $v_{1:J}$ values that are observed among the cells for which the test was completed. That way, the test will be performed using the exact distribution for the maximum possible number of cells, and the approximation will be applied only to the exact pmf components with the largest $n_j$ values, thereby keeping $\Delta$ as small as possible (Fig. 5).

## 6. Conclusions

The STA-BPRS test is accelerated by approximating symmetrical Binomial distributions by Normal distributions with the same mean and variance. The accuracy of the approximation versus the computation speed is traded off by a single control variable. The approximation makes it possible to complete the test within a practical amount of time, as opposed to centuries, for some cells. A procedure is proposed for analyzing large sets of cells to maximize the use of the exact test.

**Disclosure statement:** The author reports there are no competing interests to declare.

**Data availability statement:** Data used in this study are available at Zhang et al. (2014).

**Funding:** This research did not receive any specific grant from funding agencies in the public, commercial, or not-for-profit sectors.

**CRediT author statement:**

**Murat Okatan:** Conceptualization, Methodology, Software, Validation, Formal analysis, Resources, Writing - Original Draft, Writing - Review & Editing, Visualization, Supervision, Project administration.

# APPENDIX A

Table A.1. Breakdown of the data on the basis of multiple spike occurrences per stimulus frame.

| Cell number | $n$ | $J$ | $n_1$ | $n_2$ | $n_3$ | $n_4$ | $n_5$ | $n_6$ | $v_{1:J}$ |
|---|---|---|---|---|---|---|---|---|---|
| 20080516_R1 | | | | | | | | | |
| 1 | 38845 | 3 | 26032 | 6114 | 195 | | | | 31201591820 (3.120159e+10) |
| 2 | 4908 | 4 | 3806 | 418 | 82 | 5 | | | 794376234 (7.943762e+08) |
| 3 | 23750 | 4 | 10193 | 5984 | 523 | 5 | | | 191818866960 (1.918189e+11) |
| 4 | 13184 | 3 | 10106 | 1449 | 60 | | | | 893964150 (8.939642e+08) |
| 5 | 24896 | 3 | 19104 | 2806 | 60 | | | | 3271291835 (3.271292e+09) |
| 6 | 3496 | 5 | 1593 | 721 | 152 | 0 | 1 | | 352165608 (3.521656e+08) |
| 7 | 25717 | 3 | 22733 | 1489 | 2 | | | | 101620980 (1.016210e+08) |
| 20080516_R2 | | | | | | | | | |
| 1 | 8105 | 3 | 6161 | 963 | 6 | | | | 41581176 (4.158118e+07) |
| 2 | 11317 | 4 | 3702 | 2709 | 727 | 4 | | | 36527873200 (3.652787e+10) |
| 3 | 9955 | 5 | 3462 | 2043 | 726 | 56 | 1 | | 586641314616 (5.866413e+11) |
| 4 | 21861 | 4 | 15104 | 2850 | 315 | 28 | | | 394641749220 (3.946417e+11) |
| 5 | 1606 | 2 | 1602 | 2 | | | | | 4809 (4.809000e+03) |
| 6 | 3352 | 3 | 3289 | 30 | 1 | | | | 203980 (2.039800e+05) |
| 7 | 2569 | 3 | 1775 | 391 | 4 | | | | 3480960 (3.480960e+06) |
| 8 | 5794 | 3 | 4209 | 731 | 41 | | | | 129432240 (1.294322e+08) |
| 9 | 2571 | 4 | 603 | 499 | 286 | 28 | | | 2513546000 (2.513546e+09) |
| 10 | 16822 | 4 | 13560 | 1447 | 120 | 2 | | | 7127987064 (7.127987e+09) |
| 11 | 3967 | 4 | 1356 | 753 | 327 | 31 | | | 10739276288 (1.073928e+10) |
| 12 | 13620 | 4 | 11353 | 1029 | 67 | 2 | | | 2385702480 (2.385702e+09) |
| 13 | 9515 | 4 | 4578 | 2382 | 55 | 2 | | | 1833175176 (1.833175e+09) |
| 14 | 7376 | 3 | 4608 | 1330 | 36 | | | | 226979423 (2.269794e+08) |
| 15 | 1915 | 2 | 1749 | 83 | | | | | 147000 (1.470000e+05) |
| 16 | 6870 | 3 | 5174 | 800 | 32 | | | | 136790775 (1.367908e+08) |
| 17 | 6782 | 4 | 4642 | 1044 | 16 | 1 | | | 164965790 (1.649658e+08) |
| 18 | 2073 | 4 | 1378 | 275 | 47 | 1 | | | 36537984 (3.653798e+07) |
| 19 | 12118 | 3 | 8109 | 1961 | 29 | | | | 477354600 (4.773546e+08) |
| 20 | 1370 | 3 | 1109 | 129 | 1 | | | | 288600 (2.886000e+05) |
| 21 | 2138 | 3 | 1524 | 301 | 4 | | | | 2302750 (2.302750e+06) |
| 22 | 2774 | 3 | 1847 | 441 | 15 | | | | 13069056 (1.306906e+07) |
| 23 | 23615 | 6 | 6127 | 4334 | 2058 | 612 | 36 | 3 | 4962338604454080 (4.962339e+15) |
| 24 | 22231 | 4 | 16890 | 2557 | 73 | 2 | | | 9591993516 (9.591994e+09) |
| 25 | 14102 | 4 | 10811 | 1448 | 129 | 2 | | | 6109969320 (6.109969e+09) |
| 26 | 4877 | 5 | 1516 | 953 | 442 | 31 | 1 | | 41031524736 (4.103152e+10) |
| 27 | 14120 | 3 | 13352 | 381 | 2 | | | | 15302538 (1.530254e+07) |
| 28 | 38701 | 4 | 27222 | 5566 | 113 | 2 | | | 51830250822 (5.183025e+10) |
| 29 | 580 | 3 | 254 | 121 | 28 | | | | 902190 (9.021900e+05) |
| 30 | 19851 | 3 | 16043 | 1859 | 30 | | | | 925097040 (9.250970e+08) |
| 31 | 23777 | 3 | 20883 | 1438 | 6 | | | | 210364532 (2.103645e+08) |
| 32 | 7516 | 5 | 1608 | 1312 | 885 | 151 | 5 | | 1707062139744 (1.707062e+12) |
| 33 | 42559 | 3 | 28258 | 6921 | 153 | | | | 30123754892 (3.012375e+10) |
| 34 | 11578 | 4 | 6519 | 1847 | 439 | 12 | | | 68920051200 (6.892005e+10) |